%% file: main.tex
\documentclass[acmsmall]{acmart}

\usepackage{preprint_header}

\AtBeginDocument{%
  }

\setcopyright{none}
\usepackage[english]{babel}
\usepackage{blindtext}
\usepackage{subcaption}
\usepackage{tabularx}
\usepackage{tikz}
\usepackage{listings}
\usepackage{wrapfig}

\newcommand{\one}{({\em i}\/)\xspace}
\newcommand{\two}{({\em ii}\/)\xspace}
\newcommand{\three}{({\em iii}\/)\xspace}

\newcommand{\pb}[1]{\vspace{0.75ex}\noindent{\bf \em #1}\hspace*{.3em}}
\newcommand*\circled[1]{\tikz[baseline=(char.base)]{
            \node[fill=lightgray!20,shape=circle,draw,inner sep=2pt] (char) {\textbf #1};}}

\begin{document}

\title[Challenging Tribal Knowledge]{Challenging Tribal Knowledge -- Large Scale Measurement Campaign on Decentralized NAT Traversal}

\author{Dennis Trautwein}
\orcid{0000-0002-8567-2353}
\affiliation{%
   \institution{University of Göttingen}
  \country{Germany}
}
\email{research@dtrautwein.eu}

\author{Cornelius Ihle}
\orcid{0000-0002-3994-5218}
\affiliation{%
  \institution{University of Göttingen}
  \country{Germany}
}
\email{ihle@gipplab.org}

\author{Moritz Schubotz}
\orcid{0000-0001-7141-4997}
\affiliation{%
  \institution{FIZ Karlsruhe -- Leibniz Institute for Information Infrastructure}
  \country{Germany}
}
\email{moritz.schubotz@fiz-karlsruhe.de}

\author{Bela Gipp}
\orcid{0000-0001-6522-3019}
\affiliation{%
  \institution{University of Göttingen}
  \country{Germany}
}
\email{gipp@uni-goettingen.de}

\renewcommand{\shortauthors}{Trautwein et al.}

\include{sections/0.abstract}




\maketitle
\thispagestyle{preprintbox}

\input{sections/1.introduction}
\input{sections/2.network-address-translators}
\input{sections/3.dcutr}
\input{sections/4.measurement}
\input{sections/5.evaluation}
\input{sections/6.discusssion}
\input{sections/7.related-work}
\input{sections/8.conclusion}


\clearpage
\bibliographystyle{ACM-Reference-Format}
\bibliography{references}

\appendix
\input{sections/9.appendix}

\end{document}

%% file: sections/0.abstract.tex
\begin{abstract}

    The promise of decentralized peer-to-peer (P2P) systems is fundamentally gated by the challenge of Network Address Translation (NAT) traversal, with existing solutions often reintroducing the very centralization they seek to avoid. This paper presents the first large-scale, longitudinal measurement study of a fully decentralized NAT traversal protocol, Direct Connection Upgrade through Relay (DCUtR), within the production libp2p-based IPFS network. Drawing on over 4.4 million traversal attempts from 85,000+ distinct networks across 167 countries, we provide a definitive empirical analysis of modern P2P connectivity. We establish a contemporary baseline success rate of $70\% \pm 7.1\%$ for the hole-punching stage, providing a crucial new benchmark for the field. Critically, we empirically refute the long-held 'tribal knowledge' of UDP's superiority for NAT traversal, demonstrating that DCUtR's high-precision, RTT-based synchronization yields statistically indistinguishable success rates for both TCP and QUIC ($\sim70\%$). Our analysis further validates the protocol's design for permissionless environments by showing that success is independent of relay characteristics and that the mechanism is highly efficient, with $97.6\%$ of successful connections established on the first attempt. Building on this analysis, we propose a concrete roadmap of protocol enhancements aimed at achieving universal connectivity and contribute our complete dataset to foster further research in this domain.
\end{abstract}

%% file: sections/1.introduction.tex
\section{Introduction}
\label{sec:introduction}

The World Wide Web has become an integral part of modern life. Numerous consumer-facing services, such as Facebook, TikTok, YouTube, Netflix, and Amazon, are widely used for communication, education, entertainment, and commerce. Despite their apparent benefits, these services have come under criticism due to the potential risks of serving as single points of technical or organizational failures, becoming data monopolies, and gatekeepers~\cite{Trinh2022, bommelaer2019global}. The resulting compromised privacy, and generally, loss of self-sovereign control over one's data, poses significant risks. The vast amounts of personalized information collected by these services have put them in a position of unprecedented power to influence users' lives and even societies as a whole. However, unlike similarly influential institutions that are legitimized through democratic principles, these services lack the same legitimization~\cite{mcintosh2018we}.

In response to this, there has been a growing movement, colloquially referred to as the ``Decentralized Web''~\cite{vojir2022}. Although there is no consensus on the definition of this term, it reflects the values and ideas of challenging the power hierarchy of the current Web by eliminating intermediaries in web transactions and primarily deploying a decentralized, often peer-to-peer (P2P), architecture. However, numerous challenges arise from cutting down central service providers due to the Internet infrastructure's current adaptation to the client-server paradigm. One such challenge is the widespread deployment of Network Address Translators (NATs).


%

NAT technology, initially developed to address IPv4 address space depletion, is widely used by Internet Service Providers (ISPs) and also in corporate networks to enable multiple devices to share a single public IP address. This usually results in unobstructed access from within a local area network (LAN) to the Internet, but in combination with firewalls, it hinders incoming connections from the Internet to the LAN. This provides advantages, but also prohibits ubiquitous connectivity in peer-to-peer networks.

NAT traversal techniques such as NAT ``hole punching'' have been developed to overcome the challenge of ubiquitous P2P connectivity~\cite{Ford2005PeertoPeerCA}. In its simplest form, NAT hole punching allows two peers behind NATs to establish a direct connection with each other by simultaneously opening a connection to the public IP/Port combinations of each other. Traditional hole punching techniques like the ones used by WebRTC~\cite{rfc8825}, rely on a signaling server to facilitate the synchronization, thus inheriting all the aforementioned centralization concerns.

The decentralization of this infrastructure presents an opportunity to bolster a network's resilience against targeted attacks and censorship efforts, while simultaneously alleviating the burden of operating and maintaining such infrastructure. As a result the Direct Connection Upgrade through Relay (DCUtR) protocol~\cite{Seemann2022-jl} was designed, which draws inspiration from related protocols, including STUN, TURN, and ICE~\cite{rfc8489, rfc8656, rfc8445}. DCUtR is differentiated from its predecessors by its independence from centralized infrastructure to coordinate hole punching attempts. All involved protocols' resource requirements are minimal, making it feasible for any peer-to-peer network participant to provide these services.

In this paper, we specifically focus on the decentralized InterPlanetary File System (IPFS)~\cite{Trautwein2022,benet2014ipfs} network with its around 7k online peers\footnote{\url{https://probelab.io/ipfs/kpi/}}. 
IPFS is built upon the libp2p peer-to-peer networking stack of which the DCUtR protocol is part of and there exist around 100k~\cite{maxinden} compatible online nodes at any given time. 
To evaluate DCUtR's real-world performance, we structure our investigation around five central hypotheses (detailed in Section~\ref{sec:dcutr-hypotheses}). We seek to establish its baseline efficacy (H1), verify its independence from relay characteristics (H2), probe the efficiency and transport-agnosticism of its synchronization mechanism (H3a, H3b), and assess the practical impact of its built-in optimizations (H4). We present results from a large-scale measurement campaign designed to provide the empirical evidence needed to validate or refute these foundational claims. The contributions are as follows:


\begin{enumerate}
    \item We present the first large-scale, longitudinal measurement study of a fully decentralized NAT traversal protocol in a production P2P network. Our findings, derived from over 4.4 million data points across 85,000+ networks, establish a contemporary baseline success rate of approximately $70\%$ for the hole punching stage of direct connection establishment in the wild. This result provides a crucial, modern benchmark for the viability of P2P connectivity without centralized coordinators.
    \item We empirically challenge long-standing ``tribal knowledge'' that UDP-based protocols are inherently superior for NAT traversal. In contrast to conventional wisdom, we demonstrate that DCUtR achieves statistically indistinguishable success rates for both TCP and QUIC (which uses UDP). This finding suggests that modern, precisely synchronized hole-punching mechanisms can overcome the historical disadvantages of TCP, rendering the protocol effectively transport-agnostic.
    \item We validate DCUtR's core design, confirming the feasibility of its permissionless relaying architecture and quantifying the high efficiency of its synchronization mechanism. Our analysis shows that hole-punching success is robustly independent of the network characteristics (RTT, path location) of the randomly-selected relay peer. We also find that successful hole punches occur on the first attempt in $97.6\%$ of cases, highlighting opportunities for protocol optimization. Building on this analysis, we propose a concrete roadmap of protocol enhancements aimed at achieving universal connectivity.
    \item We contribute a unique, large-scale, and openly accessible dataset to the research community. This dataset, along with our measurement tools, is intended to foster further investigation and advancements in P2P networking, NAT traversal, and the characterization of the modern Internet's topology.
\end{enumerate}
The remainder of this paper is organized as follows. In Section~\ref{sec:network-address-translation}, we provide background information on Network Address Translation (NAT), including a taxonomy of NAT behavior and traversal techniques. Section~\ref{sec:dcutr} introduces Direct Connection Upgrade through Relay (DCUtR), a new, decentralized protocol designed to traverse NATs. In Section~\ref{sec:measurement}, we detail our measurement methodology, describe our large-scale measurement campaign, and introduce our open access dataset. Section~\ref{sec:evaluation} presents our evaluation and analysis, followed by a discussion of the measurement results in Section~\ref{sec:discussion}. Section~\ref{sec:optimizations} introduces a roadmap for protocol enhancements and Section~\ref{sec:related-work} reviews related work in the field. Finally, Section~\ref{sec:conclusion} concludes the paper.


%% file: sections/2.network-address-translators.tex
\section{The Challenge of NAT for Peer-to-Peer Connectivity}
\label{sec:network-address-translation}


Network Address Translation (NAT) is a prevalent technique used in network gateways to modify IP address information, primarily to mitigate IPv4 address exhaustion~\cite{rfc1631}. The most common variant, Network Address Port Translation (NAPT), also known as Port Address Translation (PAT), allows multiple devices within a private network to share a single public IP address~\cite{rfc2663}. The NAPT device maintains a dynamic translation table, mapping an internal source `(IP:port)` tuple to a public-facing `(IP:port)` tuple for each outbound connection.

This mapping and filtering mechanism allows outgoing connections but blocks unsolicited inbound traffic, which obstructs peer-to-peer (P2P) networking. A gateway only forwards an incoming packet if it matches a recent outgoing connection; otherwise, the packet is dropped. Hence, this mechanism requires specialized NAT traversal techniques to establish direct P2P links.


\subsection{Taxonomy of NAT Behavior}

The viability of any NAT traversal technique is dictated by the specific mapping and filtering behaviors of the NAT device~\cite{rfc5382}, as standardized in RFC 4787~\cite{rfc4787}. Rather than treating these properties in isolation, it is more instructive to examine how they combine to form distinct NAT archetypes, which directly impact P2P compatibility.

The most P2P-friendly NATs are known as \textbf{Cone NATs}. Their defining characteristic is \textit{Endpoint-Independent Mapping (EIM)}, where the NAT assigns a stable public endpoint `(IP:port)` to an internal endpoint, reusing this mapping for all subsequent outbound connections, regardless of their destination. This predictability is the cornerstone of hole punching, as a peer can learn its public endpoint and reliably share it with others. Cone NATs are further distinguished by their filtering policies: \textit{Full Cone NATs} (Endpoint-Independent Filtering) are the most permissive, allowing inbound traffic from any source once a mapping exists. \textit{Restricted Cone NATs} (Address-Dependent or Address and Port-Dependent Filtering) impose stricter rules, only permitting inbound packets from an external endpoint `(IP:port)` to which the internal peer has recently sent traffic.

In stark contrast, a \textbf{Symmetric NAT} presents the most significant challenge to P2P connectivity. It employs \textit{Address and Port-Dependent Mapping (APDM)}, creating a new, unique public mapping for each distinct destination endpoint. Consequently, the public endpoint a peer observes when contacting a discovery server is different from the one its NAT would use for a connection to another peer. This unpredictability renders simple hole punching ineffective, as the shared address information becomes invalid for establishing a new connection. This behavior, combined with restrictive filtering, often makes a relay-based solution the only viable option for establishing connectivity.

\subsection{NAT Traversal Techniques}

The de-facto standard for navigating the complexities of NAT is the Interactive Connectivity Establishment (ICE) framework~\cite{rfc5245, rfc8825}, which orchestrates several techniques to find the most efficient path between peers. The ICE process systematically attempts to establish connectivity using:

\begin{enumerate}
    \item \textbf{Address Discovery with STUN:} Peers first query a centralized \textit{Session Traversal Utilities for NAT (STUN)} server~\cite{rfc8489}. The STUN server reflects the peer's public IP address and port as seen from the public internet (its "server reflexive" address) and helps classify the NAT's behavior.

    \item \textbf{Direct Connection via Hole Punching:} Armed with their respective public addresses, peers exchange this information through a signaling channel (often centrally operated) and attempt to establish a direct connection via \textit{hole punching}~\cite{rfc5128}. This involves sending simultaneous connection-request packets to each other, exploiting the temporary mappings created in their NATs by their own outbound traffic.

    \item \textbf{Relay Fallback with TURN:} If direct connection attempts fail—typically when one or both peers are behind a Symmetric NAT—ICE falls back to its last resort: relaying traffic through a \textit{Traversal Using Relays around NAT (TURN)} server~\cite{rfc8656}. While this guarantees connectivity, it comes at the cost of increased latency and requires the TURN server to handle the bandwidth for the entire session.
\end{enumerate}

\noindent This NAT traversal technique aims to overcome the barrier that NATs impose on direct communication, offering a range of solutions from ``best-effort'' (hole punching) to guaranteed connectivity (TURN relaying), with ICE being the orchestrator that tries to find the most efficient path.

Other mechanisms, such as UPnP~\cite{rfc6970} for automatic port forwarding or Application-Level Gateways (ALGs)~\cite{rfc2663}, exist but are not universally reliable or are often disabled due to security concerns\footnote{https://www.cisa.gov/news-events/news/home-network-security}. Therefore, the robustness of the modern P2P ecosystem, particularly in applications like WebRTC, heavily relies on the availability of centralized STUN, TURN, and signaling infrastructure. This dependency introduces operational costs and single points of failure, running counter to the principles of fully decentralized systems, a limitation that motivates the development and analysis of protocols such as DCUtR.

%% file: sections/3.dcutr.tex
\section{The libp2p DCUtR Protocol}
\label{sec:dcutr}

The Decentralized Connection Upgrade Through Relay (DCUtR) protocol is based on \texttt{libp2p}\footnote{\url{https://libp2p.io}}, a modular peer-to-peer networking stack in which DCUtR is implemented. \texttt{libp2p} is used by major peer-to-peer networks like Ethereum~\cite{Buterin2013} and IPFS~\cite{Trautwein2022,benet2014ipfs} as it provides foundational tools for peer discovery, connection establishment, stream multiplexing, and secure communication. Within libp2p, several precursor protocols play a crucial role in gathering the necessary information and establishing initial connectivity for hole punching. These are notably the \textbf{Identify}\footnote{\url{https://docs.libp2p.io/concepts/fundamentals/protocols/\#identify}}, \textbf{AutoNAT}\footnote{\url{https://docs.libp2p.io/concepts/nat/autonat/}}, and the \textbf{Circuit v2}\footnote{\url{https://docs.libp2p.io/concepts/nat/circuit-relay/}} relay protocol. To be able to understand the following measurement methodology and evaluation, we give a brief overview of these precursor protocols as well as the DCUtR protocol operations. We point the reader to~\cite{Seemann2022-jl} for a more comprehensive description of the DCUtR protocol.

\subsection{libp2p}
\label{sec:dcutr-libp2p}

Identify, AutoNAT, and Circuit v2 protocols allow peers to discover their external network addresses and reachability. Further, the protocols establish preliminary connections without sole reliance on centralized STUN and TURN servers.

\subsubsection{Identify}

The Identify protocol provides functionality similar to STUN, is lightweight, and all libp2p peers support it by default. Upon connection (also relayed ones), peers exchange messages through which a peer learns its own externally observed public IP address(es) and port from its communication partner's perspective. This leverages existing connections, negating the need for separate, centralized STUN infrastructure and potentially yielding more reliable address information as it uses the same transport protocols as the eventual direct connection.

\subsubsection{AutoNAT}
\label{sec:dcutr-autonat}
Complementing this, the AutoNAT protocol determines a peer's actual reachability at these observed addresses. A peer requests others' to attempt dialing back to its advertised addresses. A successful dial-back classifies the peer as public, while failure indicates a private peer that likely cannot receive unsolicited incoming connections. On top of AutoNAT, a libp2p host tries to establish port mappings using the Universal Plug and Play (UPnP) and/or Port-Mapping Protocol (PMP). However, this is purely advisory as \one routers may not reliably report the mapping's status or \two may silently expire the mapping at any time. AutoNAT will authoritatively inform the libp2p host over which addresses it is dialable from the outside.

\subsubsection{Circuit v2}
\label{sec:dcutr-circuit-v2}
For private peers, the Circuit v2 protocol provides relaying services, primarily for light signaling and coordination purposes, such as for DCUtR. Crucially, Circuit v2 is designed to require minimal resources, imposing negligible processing or bandwidth costs on the relays. This contrasts with traditional TURN relays that handle entire traffic streams. Circuit v2 achieves this by requiring private peers to obtain reservations from relays, which are limited in number, duration, and data volume for any single relayed connection. This design allows the vast majority of public libp2p peers to act as relays without significant resource burden. The establishment of this initial relayed connection via Circuit v2 is the precursor from which DCUtR aims to upgrade to a direct connection.

\subsection{The DCUtR Protocol}
\label{sec:dcutr-protocol}

\begin{figure}
    \centering
    \includegraphics[width=.8\linewidth]{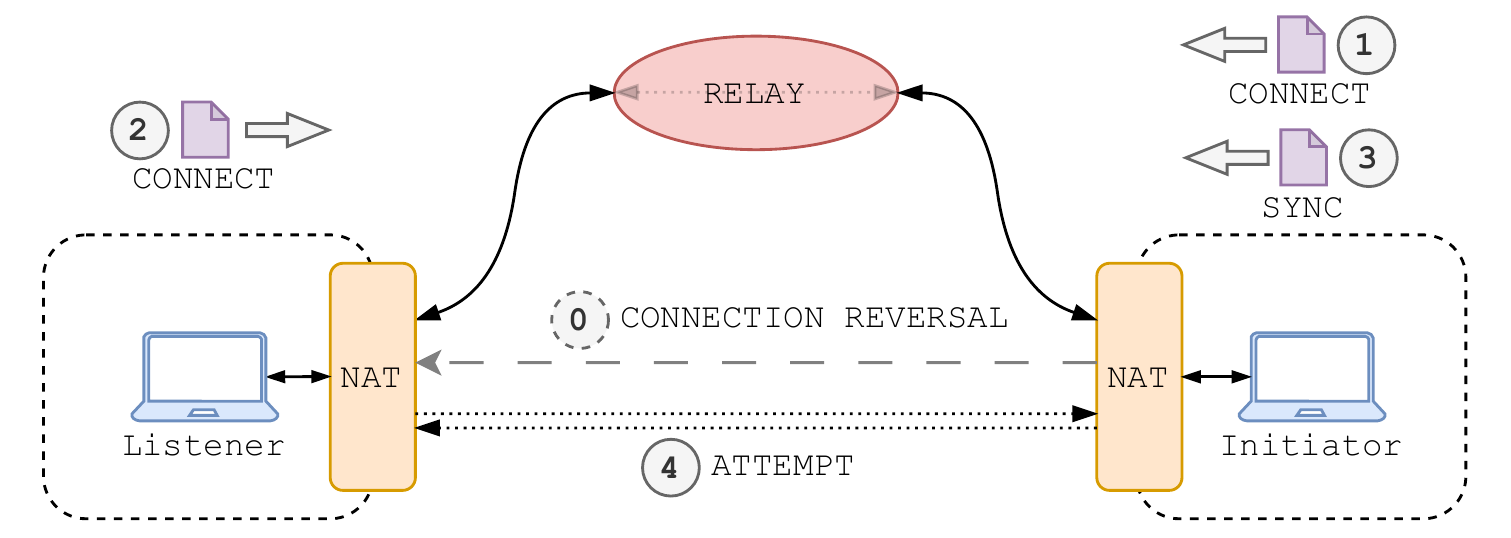}
    \caption{DCUtR protocol flow diagram}
    \label{fig:dcutr-flow-chart}
\end{figure}

The DCUtR protocol attempts to upgrade an existing relayed libp2p connection to a direct one. The process involves an \textbf{initiator}, the peer who accepted the relayed connection, and a \textbf{listener} who waits for the other party to initiate the DCUtR protocol exchange. Figure~\ref{fig:dcutr-flow-chart} depicts the protocol exchange which we will reference in the following. The whole setup starts with the initiator having a valid reservation at any relay in the network. Then the listener can use the relay to establish a ``limited'' connection through the relay to start the exchange of identify and DCUtR protocol messages. First, the initiator may bypass the whole hole punching procedure and attempt a ``Connection Reversal''. If the listener appears public, based on the Identify data that was exchanged over the relayed connection, the initiator directly dials its advertised addresses~\circled{0}. If successful, a direct connection is formed without needing additional hole punching. If Connection Reversal fails or the listener is private, the full hole punching procedure begins:

\begin{enumerate}
    \item \textbf{Address Exchange:} The initiator sends a \texttt{CONNECT} message~\circled{1} containing its candidate public addresses (from Identify) to the listener via the relay. The listener responds with its own \texttt{CONNECT} message~\circled{2} containing its addresses. The initiator uses this exchange to measure the round-trip time (RTT) of the relayed path.
    \item \textbf{Synchronization:} The initiator then sends a SYNC message~\circled{3} to the listener over the relay and waits half the measured RTT for the next step.
    \item \textbf{Dialing:} Upon receiving \texttt{SYNC}, the listener immediately attempts to establish a direct connection to the initiator's advertised addresses. The initiator, having waited half the measured RTT, also dials the listener's addresses~\circled{4}. This synchronized attempt aims for simultaneous packet arrivals at both NATs, creating the necessary mappings for a direct connection.
\end{enumerate}

\noindent This process uses transport-specific mechanisms like TCP Simultaneous Open or a QUIC-specific technique involving one side acting as client and the other sending dummy UDP packets to establish NAT state. The synchronization and synchronized dialing will be retried twice if a direct connection couldn't be established. Once a direct connection is established, the relayed connection is closed. The DCUtR coordination itself is designed to be lightweight, typically involving two network round-trips and exchanging less than 500 bytes per direction over the relay. A detailed sequence diagram is in Appendix~\ref{appendix:dcutr-sequence-diagram}. As mentioned earlier, for a more detailed discourse and a comprehensive description, we point the reader to Seemann et al.~\cite{Seemann2022-jl}.

\subsection{Protocol Hypotheses}
\label{sec:dcutr-hypotheses}

The design of DCUtR, as a decentralized protocol intended for large-scale, permissionless P2P networks, gives rise to a set of foundational questions regarding its real-world performance and behavior. To structure our investigation, we formulate these questions as testable hypotheses. Each hypothesis stems directly from a core design goal of the protocol or a known challenge in NAT traversal. The measurement campaign detailed in Section~\ref{sec:measurement} gathers the empirical evidence needed to validate or refute these central claims.


\pb{Viability and Efficacy.} The most fundamental question for any new protocol is its basic efficacy. The public internet presents a hostile environment for P2P connectivity, with a share of NAT devices (e.g., Symmetric NATs) designed to be exceptionally difficult to traverse~\cite{acunto2009}. For DCUtR to be considered a viable connection strategy, we establish a conservative but critical benchmark for viability at 50\% and require that it must succeed more often than it fails.
\begin{description}
    \item[H1:] We hypothesize that DCUtR can achieve a success rate greater than 50\% for the hole punching stage, demonstrating its fundamental viability as a robust and effective decentralized NAT traversal solution in a real-world, heterogeneous environment. Evaluated in Section~\ref{sec:analysis-efficacy}.
\end{description}

\pb{Relay Independence.} A core tenet of the protocol is its decentralized nature, allowing any public peer to act as a signaling relay. This architecture is only robust if the protocol's success is not dependent on a small set of privileged, high-performance, or strategically placed relays.
\begin{description}
    \item[H2:] We hypothesize that the success of a DCUtR hole punch is largely independent of the network characteristics of the facilitating relay. This would validate the architectural choice of allowing any public peer to act as a relay in a permissionless network. Evaluated in Section~\ref{sec:analysis-relay-independence}.
\end{description}

\pb{Effective Synchronization.} The protocol's design relies on a precise RTT-based synchronization mechanism to coordinate the hole punch. A highly effective synchronization mechanism should lead to two key outcomes: efficiency and transport-agnosticism. First, it should be so precise that the hole punch succeeds on the first try, making subsequent retries largely unnecessary. Second, this precision should mitigate the timing challenges that have historically made TCP hole punching significantly harder than UDP-based methods. This leads to two distinct but related hypotheses (evaluated in Section~\ref{sec:effective-synchronization}):

\begin{description}
    \item[H3a:] We hypothesize that the DCUtR synchronization mechanism is highly efficient, with successful traversals predominantly occurring in the initial attempt.
    \item[H3b:] We hypothesize that DCUtR's performance is not fundamentally tied to the underlying transport protocol, achieving comparable success rates for both TCP and QUIC and challenging the conventional ``tribal knowledge'' of UDP's inherent superiority.
\end{description}

\pb{Optimization Effectiveness.} The protocol explicitly includes ``Connection Reversal'' as a fast-path optimization to bypass the full hole-punching procedure when a peer has an existing port mapping (e.g., via UPnP). Any evaluation must verify if such documented optimizations are effective in practice.
\begin{description}
    \item[H4:] We hypothesize that the Connection Reversal mechanism is an effective optimization, predicting it will significantly increase the likelihood of direct connections for peers with favorable NAT configurations, thereby avoiding the overhead of the full hole-punching exchange. Evaluated in Section~\ref{sec:optimization-effectiveness}.
\end{description}

%% file: sections/4.measurement.tex
\section{Measurement Methodology \& Campaign}
\label{sec:measurement}

The objective of our measurements is to assess the performance of the DCUtR protocol, with a specific focus on validating or refuting the hypotheses in Section~\ref{sec:dcutr-hypotheses}. Our goal is to gather insights of the protocols efficacy and into areas for protocol improvement. To achieve these objectives, we conducted a measurement campaign designed to comprehensively assess the performance of the DCUtR protocol. This section covers the measurement architecture, details on our measurement campaign, and dataset of our research.

\subsection{Methodology}
\label{sec:methodology}

\begin{figure}[t]
    \centering
    \includegraphics[width=.8\linewidth]{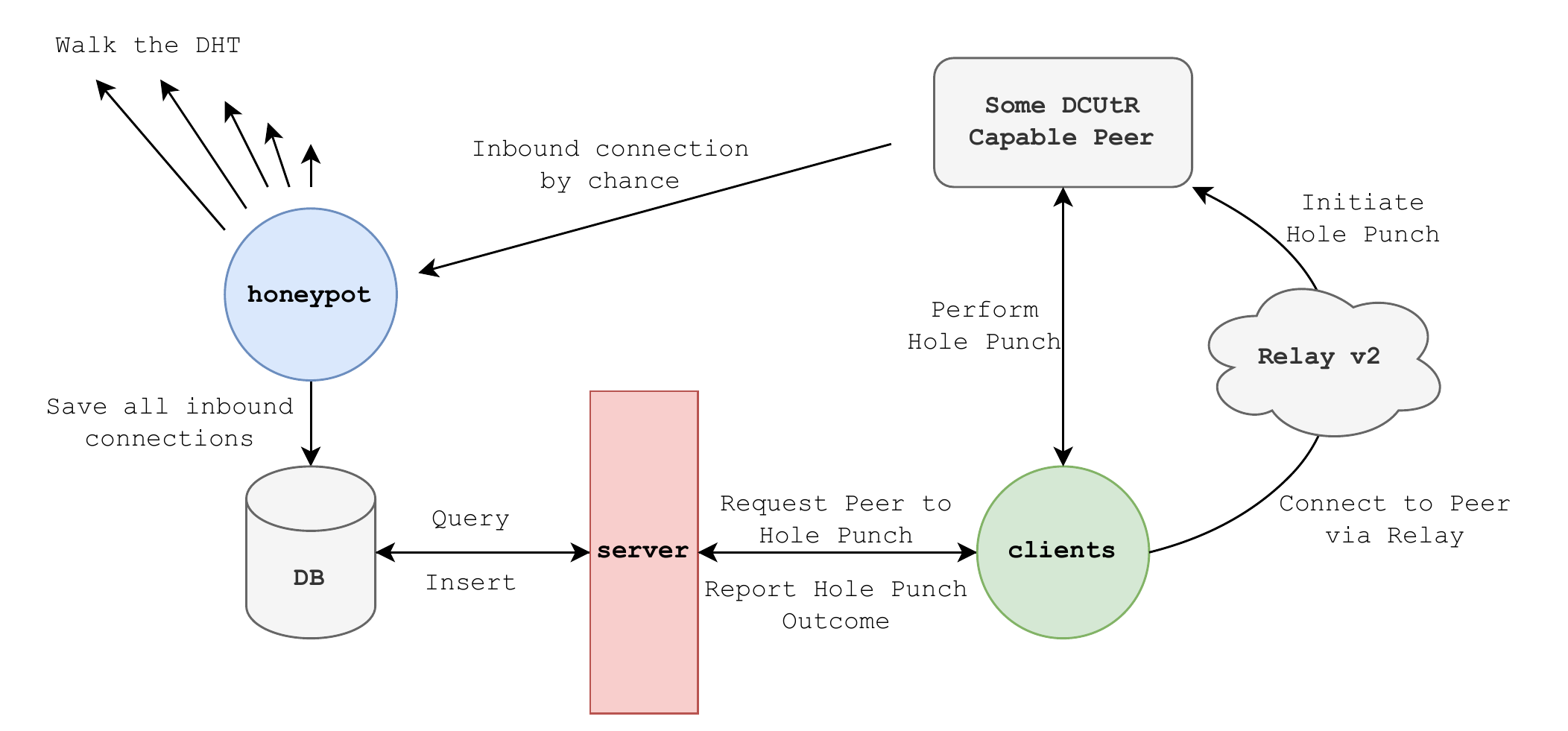}
    \caption{Measurement infrastructure architecture. The central components \textbf{honeypot}, \textbf{server}, and \textbf{clients} allow us to detect, serve and hole punch DCUtR-capable remote peers.}
    \label{fig:architecture}
\end{figure}

The main challenge in evaluating the performance of the DCUtR protocol is the discovery of DCUtR-capable peers. For a meaningful assessment of NAT traversal capabilities, target peers must be situated behind NATs, rendering them not publicly reachable. This inherent characteristic, however, complicates their discovery within decentralized and permissionless peer-to-peer networks as they typically, and specifically in the case of the IPFS network, lack a central registry of participants or their supported protocols.

Standard peer discovery mechanisms, such as querying public distributed hash tables (DHTs)~\cite{Trautwein2022}, are insufficient for this purpose. By default, only publicly reachable peers announce their presence to the DHT, thereby excluding the very NAT'ed peers crucial for our measurement campaign. To address this challenge, we introduce the \textbf{honeypot} component.

\subsubsection{Honeypot}

The honeypot is a DHT server peer that is designed to be very stable and announces itself to the network by slowly crawling it~\cite{Trautwein2022}. This increases the chances that other peers add the honeypot peer to their routing tables. This, in turn, leads to increased traffic from client peers to the honeypot because other server peers increasingly redirect clients to it. The honeypot tracks all inbound connections from peers that satisfy both of the following conditions: \one peer supports the DCUtR protocol \two peer is only reachable through relay addresses which indicates that it is behind a NAT. ``Tracking all inbound connections'' means that they are stored in the common database that is then queried and served by the \textbf{server} component.

\subsubsection{Server}

The server component exposes a Google Remote Procedure Call (gRPC) API allowing clients to query for NAT'd + DCUtR capable peers that were previously discovered by the honeypot. It is also responsible for tracking the results of hole punch attempts and provides a centralized location for the clients to retrieve information about potential peers to hole punch. A key element of our methodology is the ``protocol filter'', a mechanism that allows us to isolate and evaluate the performance of specific transport protocols (TCP and QUIC) independently. When a client queries the server for a peer to hole punch the server randomly assigns a ``protocol filter'' which the client should apply to the hole punch. A protocol filter of ``TCP'' would mean the client will \one only announce the public TCP address it is listening on to the remote peer over the relay and \two only use the public TCP Multiaddress of the remote peer for the hole punch. That way both sides only utilize the respective protocol despite potentially listening on additional ones.

\subsubsection{Clients}

The clients periodically query the server for peers to probe. They then use this information to perform hole punching attempts and report back the outcomes to the server. The clients connect to the peer via their advertised relayed addresses acting as the previously introduced \textit{listener}. They then wait for the remote peer, \textit{initiator}, to initiate the hole punch protocol. This component provides the actual data for the measurement campaign and allows the server to collect the results. The client is written in Go and comes in two variants, a CLI and a GUI version. The GUI version was developed to ease the onboarding of users to the measurement campaign. The source code is openly accessible under \href{https://github.com/REDACTED}{https://github.com/REDACTED}.

\subsubsection{Interplay}

Figure~\ref{fig:architecture} depicts the interplay of the different components. It shows the honeypot that slowly crawls or ``walks'' the DHT to increase chances that DCUtR capable but NAT'd peers connect to it. These inbound connections are stored in a database which is then queried and served to clients. The clients use the served relay address(es) to connect to the remote peer and trigger the DCUtR protocol. After a direct connection was successfully established, three DCUtR attempts have failed, or connection establishment to the remote peer timed out, the client reports back the result to the server which then persists the data for further analysis.


\subsection{Campaign}
\label{sec:campaign}
\begin{figure*}[t!]
    \centering
    \begin{subfigure}[t]{0.48\textwidth}
        \centering
        \includegraphics[width=\linewidth]{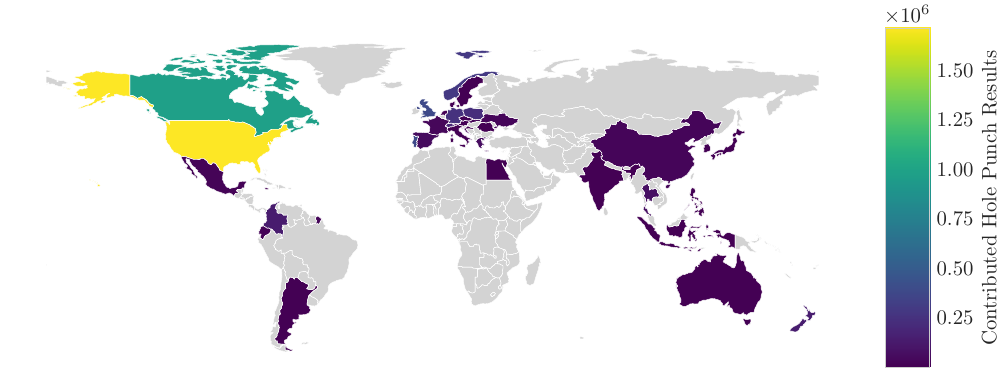}
        \caption{}
        \label{fig:geo-clients}
    \end{subfigure}%
    ~ 
    \begin{subfigure}[t]{0.48\textwidth}
        \centering
        \includegraphics[width=\linewidth]{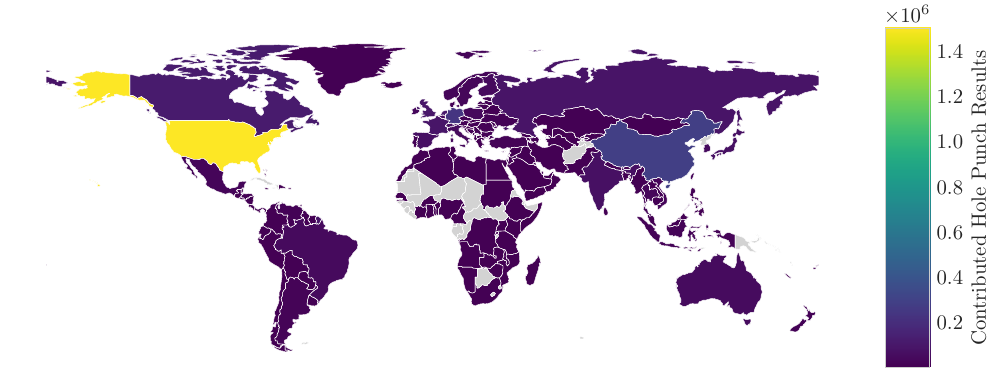}
        \caption{}
        \label{fig:geo-remote-peers}
    \end{subfigure}
    \caption{(a) Geographic distribution of controlled client peers in the measurement campaign that contributed hole punch results. (b) Geographic distribution of remote peers that interacted with the IPFS network that contributed hole punch results in the measurement campaign.}
\end{figure*}

The measurement campaign officially ran from the 1st of December 2022 until the 1st of January 2023. We announced the campaign to public channels in the libp2p and IPFS communities as well as internally within REDACTED. We asked interested parties to sign up with a form where we asked general questions about the network they primarily plan to run the client in. The users then received and API-Key from us so that we can link the data points to the responses in our questionnaire. However, it's important to note that we didn't impose any restrictions with regard to client mobility. Participants were free to move the client around and change networks at their discretion. This becomes a challenge because data points of a particular client need to be split by the individual networks that it ran in. We discuss this challenge of the data analysis in Section~\ref{sec:network-identification} but also believe that this measurement setup was crucial to drive participation.

Further, participation did not require sign-up through our form. People could just download the source code and run the clients themselves and report results to the server. In this case, the server generates a random API-Key for these participants. If we detected abuse by these non-signed-up clients we would be able to exclude their data points. We did not detect abuse.

\subsection{Dataset}
\label{sec:dataset}

\begin{table}
    \centering
    \caption{The possible outcomes of an individual hole punch result as reported by the clients. Each hole punch results can consists of up to three ``attempt'' data points. Their outcomes are listed in table \ref{tab:hpa-outcomes}}
    \label{tab:hpr-outcomes}
    \vspace{-0.3cm}
    \begin{tabularx}{\textwidth}{lX}
    \hline
    \hline
         Outcome & Description \\ \hline
         \textbf{\texttt{UNKNOWN}} & There is no information why and how the hole punch completed. \\
         \textbf{\texttt{NO\_CONNECTION}} & The client could not connect to the remote peer via any of the provided multi addresses. \\
         \textbf{\texttt{NO\_STREAM}} & The client could connect to the remote peer via any of the provided multi addresses but no \texttt{/libp2p/dcutr} stream was opened within 15s. That stream is necessary to perform the hole punch. \\
         \textbf{\texttt{CONNECTION\_REVERSED}} & The client only used one or more relay multi addresses to connect to the remote peer, the \texttt{/libp2p/dcutr} stream was not opened within 15s, and we still end up with a direct connection. This means the remote peer successfully reversed it. \\
         \textbf{\texttt{CANCELLED}} & The user stopped the client (also returned by the rust client for quic multi addresses). \\
         \textbf{\texttt{FAILED}} & The hole punch was attempted multiple times but none succeeded OR the \texttt{/libp2p/dcutr} was opened but we have not received the internal start event OR there was a general protocol error. \\
         \textbf{\texttt{SUCCESS}} & Any of the hole punch attempts succeeded. \\
    \hline
    \hline
    \end{tabularx}
\end{table}

\begin{table}
    \centering
    \caption{The possible outcomes of an individual hole punch attempt as reported by the clients. Each hole punch results can consists of up to three ``attempts'' data points with the possible outcomes listed below. The outcome of the attempts informs the overall outcome of the result in table~\ref{tab:hpr-outcomes}}
    \label{tab:hpa-outcomes}
    \vspace{-0.3cm}
    \begin{tabularx}{\textwidth}{lX}
    \hline
    \hline
         Outcome & Description \\ \hline
         \textbf{\texttt{UNKNOWN}} & There was no information why and how the hole punch attempt completed. \\
         \textbf{\texttt{DIRECT\_DIAL}} & The connection reversal from our side succeeded (should never happen). \\
         \textbf{\texttt{PROTOCOL\_ERROR}} & This can happen if e.g., the stream was reset mid-flight. \\
         \textbf{\texttt{CANCELLED}} & The user stopped the client. \\
         \textbf{\texttt{TIMEOUT}} & We waited for the internal start event for 15s but timed out. \\
         \textbf{\texttt{FAILED}} & We exchanged \texttt{CONNECT} and \texttt{SYNC} messages on the \texttt{/libp2p/dcutr} stream but the final direct connection attempt failed; the hole punch was unsuccessful. \\
         \textbf{\texttt{SUCCESS}} & We were able to directly connect to the remote peer.  \\
    \hline
    \hline
    \end{tabularx}
\end{table}

In total, we tracked over 6.25M hole punches from 212 API keys while registering 148 sign-ups. Figures \ref{fig:geo-clients} and \ref{fig:geo-remote-peers} show that the clients were deployed in 39 different countries and hole punched remote peers in 167 different countries. The maps also show the number of data points that were contributed from each country. Both graphs prove that even though the sample size is unevenly distributed with a focus on the US, we have globally distributed data points.

Each time a client has completed a hole punch probe, it reports back a hole punch \textbf{result} which can contain multiple hole punch \textbf{attempts} because, as mentioned previously, a hole punch is tried at most three times. Each attempt has its own ``outcome'' as well as the overall result. Tables~\ref{tab:hpr-outcomes} and~\ref{tab:hpa-outcomes} list the possible outcomes for a hole punch \textit{result} and \textit{attempt} respectively. An intuition for the different outcomes becomes relevant in the evaluation below. Further, each data point tracks the clients' and remotes' peer identifier, all IPs and ports that the client is listening on, the set of addresses that were used to connect to the remote peer (usually just a single relay address), the set of addresses used to directly connect to the peer, open connections to the remote peer after the hole punch, and potential active port mappings on the clients' side. We make our dataset available under the CC BY-SA license with this IPFS content identifier:

\begin{center}
    \href{https://bafybeia7sq3nfd7c4obcy7ahjvnoka7ujdiob33r7rqyeycgicdt3iknki.ipfs.dweb.link/}{\texttt{bafybeia7sq3nfd7c4obcy7ahjvnoka7ujdiob33r7rqyeycgicdt3iknki}}
\end{center}

\noindent A detailed description of the datasets' structure and content is in Appendix~\ref{appendix:dataset}. All personally identifiable information is anonymized.

\subsubsection{Limitations}
\label{sec:dataset:limitations}

We acknowledge methodological limitations that may influences interpretation of the evaluation.

\pb{Client Sampling.} Our client fleet was recruited from volunteers within the IPFS and libp2p communities. This group may be more technically proficient and operate on more stable or less restrictive networks than the general population of P2P users. This potential bias could lead to an overestimation of the protocol's success rate (H1) and may affect the generalizability of our findings.

\pb{Discovery Bias.} Our 'honeypot' discovers remote, NAT'd peers by accepting their inbound relayed connections. This method systematically excludes peers behind highly restrictive NATs or firewalls that prevent them from establishing even a relayed connection. Consequently, our study population of remote peers is pre-filtered for a baseline level of reachability, which also may inflate the measured hole-punching success rate.

\subsection{Ethics}
\label{sec:ethics}

The research followed strict ethical guidelines. Participants either signed up, receiving detailed information and providing explicit consent regarding data collection (including IP addresses and libp2p PeerIDs) and its academic use, or voluntarily ran a tool whose documentation (README) clearly stated data collection (including IP addresses and libp2p PeerIDs) for research.

Initially, IP addresses were stored in clear text for network analysis (e.g., geographic distribution, network identification for the DCUtR protocol) and securely on a restricted-access research server, accessible only to the core researcher. After core analysis for this paper, all IP addresses were irreversibly anonymized or deleted, preventing deanonymization while allowing statistical correlation.

Data from remote, non-consenting peers was collected based on their operation of public, permissionless P2P nodes (IPFS, libp2p), which inherently expose network metadata (IP, PeerID, protocols) through standard interactions. The ``honeypot'' acted as a regular libp2p node, observing public addresses and sending standard protocol messages. Collected data (e.g., hole punch success) was a byproduct of these standard interactions, with no non-standard or privacy-invasive data gathered beyond what a public P2P node inherently exposes. 

%% file: sections/5.evaluation.tex
\section{Evaluation}
\label{sec:evaluation}


In this Section, we evaluate the data collected during our measurement campaign to empirically test the hypotheses outlined in Section~\ref{sec:dcutr-hypotheses}. We prepare the data, analyze the protocol's overall success rate, its dependence on network factors and transport protocols, and the efficiency of its mechanisms.




\subsection{Data Preparation -- Network Identification}
\label{sec:network-identification}

The analysis of the DCUtR protocol's performance begins with the identification of individual networks from which clients operated. Due to client mobility, it is crucial to segment a single client's data points by the network from which they originated. This prevents confounding results from distinct network environments (e.g., different NAT devices and policies). The methodology for discerning individual networks relies on two assumptions: \one data points belong to the same network if the client reports the same public IP address, and \two if a client changed its public IP but remained in the same Autonomous System (AS), it is considered the same network if the set of locally assigned private IP addresses remained constant. With this technique, we discover $859$ distinct networks that the clients operate in (no two clients operated from the same network) and $86,769$ networks for remote peers.

\begin{wrapfigure}{r}{0.5\textwidth}
    \centering
    \begin{subfigure}[t]{0.21\textwidth}
        \centering
        \includegraphics[width=\linewidth]{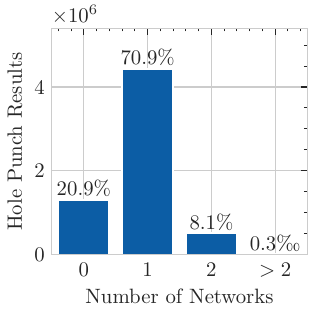}
        \caption{}
        \label{fig:networks-distribution}
    \end{subfigure}%
    \hspace{0.5cm}
    \begin{subfigure}[t]{0.20\textwidth}
        \centering
        \includegraphics[width=\linewidth]{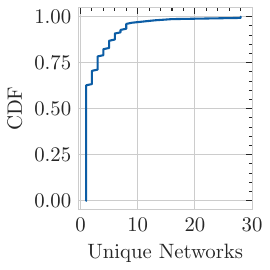}
        \caption{}
        \label{fig:client-networks}
    \end{subfigure}
    \caption{(a) Network identifications per hole punch result. Most could be linked to a single network based on the reported IP addresses that the client listens on. (b) CDF of the number of unique networks per client, showing that over $60\%$ of clients operated from a single network throughout the campaign.}
\end{wrapfigure}

Figure~\ref{fig:networks-distribution} shows that the network identification process revealed that approximately 4.43M or $70.9\%$ of the 6.25M reported hole punch data points could be unambiguously associated to a single client network. However, $20.9\%$  of data points had ``0 public networks,'' indicating cases where no public IP was advertised. This can happen if no relay reservation was in place yet or Identify was not able to determine a public address. Clients that announce public IPs in different networks account for $8.1\%$ of all reported hole punch results. 
This can happen if a client was assigned more than one public IP address.
Figure~\ref{fig:client-networks} shows that the majority of over $60\%$ of clients operated from within a single network while a single client moved between 28 different networks during the measurement campaign.

For the following analysis, we only consider hole punch results that we were able to unambiguously associate to a single network.
The excluded data points represent fundamental failures of the broader connectivity stack upon which DCUtR depends. While our subsequent analysis focuses on the efficacy of the hole-punching stage itself, this initial failure rate underscores the multi-faceted challenges of establishing P2P connectivity in the wild.
We discuss this and other limitations in Section~\ref{sec:discussion}.

\subsection{Analysis}
\label{sec:analysis}

In the following sections we evaluate the hypotheses from Section~\ref{sec:dcutr-hypotheses}.





\subsubsection{H1: Viability and Efficacy.}
\label{sec:analysis-efficacy}

\begin{figure*}[t!]
    \centering
    \begin{subfigure}[t]{0.5\textwidth}
        \centering
        \includegraphics[width=\linewidth]{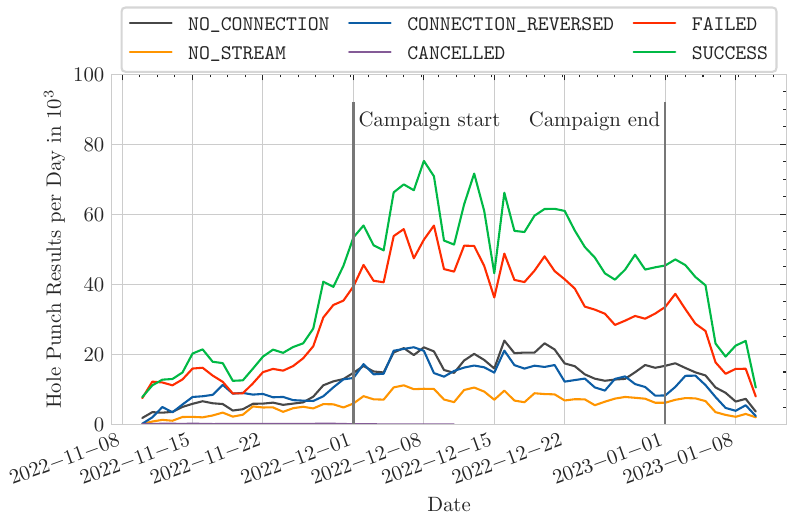}
        \caption{}
        \label{fig:campaign-outcomes}
    \end{subfigure}%
    ~ 
    \begin{subfigure}[t]{0.5\textwidth}
        \centering

    \includegraphics[width=\linewidth]{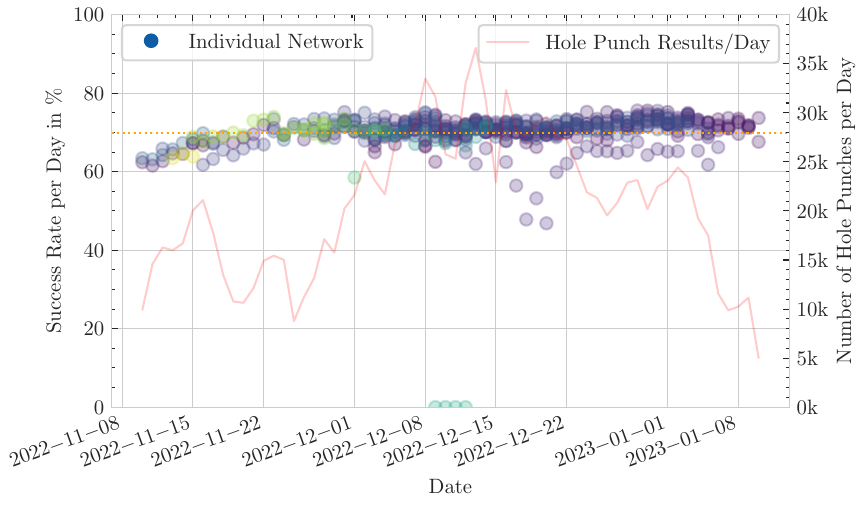}
        \caption{}
        \label{fig:network-success-rates}
    \end{subfigure}
    \caption{(a) Reported hole punch results over the course of the duration of our measurement campaign split by their individual outcomes according to table~\ref{tab:hpr-outcomes}. (b) Daily success rates of hole punches for individual networks across the entire measurement period. The dashed orange line is the line of best fit across all success rates.}
\end{figure*}

To test our primary hypothesis regarding the viability of decentralized NAT traversal (H1), we first establish the baseline success rate of the DCUtR protocol across our large-scale deployment. Figure~\ref{fig:campaign-outcomes} shows the daily number of reported hole punch results grouped by their eventual outcome. We have described the possible outcomes earlier in Table~\ref{tab:hpr-outcomes}. The graph shows that the number of contributed hole punches rose following the 1st of December and then receded over Christmas until it died away after we sent out the measurement campaign termination notification in the beginning of January. This behavior correlates and coincides with the start and end of the measurement campaign as well as events like holiday period for the majority of clients (compare Figure~\ref{fig:geo-clients} and~\ref{fig:geo-remote-peers}).

In order to derive the number for the success rate that a peer would experience when trying to connect to a random peer in the IPFS network we only take into account the hole punch result data points \one that do not have a port mapping in place \two where the client contributed more than 1k data points and \three that have an outcome of \texttt{SUCCESS} or \texttt{FAILED} (see Table~\ref{tab:hpr-outcomes}).

Figure~\ref{fig:network-success-rates} shows the success rates for each individual network as differently shaded points averaged per day. The second y-axis shows the residual number of data points given the above filtering criteria to get a sense of the sample size. Calculating the line of best fit across all networks for the measurement period yields an average 
\begin{center}
\textbf{success rate of $70\% \pm 7.1\%$ for the hole punching stage.}
\end{center}

\noindent This observed success rate across a diverse and global set of networks provides strong evidence supporting H1, confirming that DCUtR is an effective protocol for establishing direct P2P connections in the wild. We discuss this rate later in Section~\ref{sec:discussion}.



Having established the viability of the DCUtR protocol with a high success rate (H1), we now analyze in Figure~\ref{fig:latency-impact} the tangible performance benefit for peers that successfully establish a direct connection. This analysis quantifies the latency reduction achieved by upgrading from a relayed to a direct path. In this graph, we divide the RTT after a successful hole punch to the remote peer by the RTT through the relay to the remote peer. E.g., if one has an RTT to the remote of 1s through a relay and afterwards a direct RTT of 0.7s we only have $70\%$ of the original RTT. The above graph shows that $50\%$ of all peers experienced $70\%$ of their original round trip time or less after the hole punch. $10\%$ report a higher RTT after the hole punch (all points above $10^0$). Generally, $90\%$ of peers profit from a hole punch with a lower RTT confirming that hole punching can be a crucial technique to allow for a broader range of delay-sensitive applications. A direct path being slower than a relayed one is possible if the relay is located on a high-speed internet backbone, while the direct peer-to-peer path traverses multiple, slower, consumer-grade ISP networks.

\subsubsection{H2: Relay Independence}
\label{sec:analysis-relay-independence}

\begin{figure*}[t]
    \centering
    \begin{subfigure}[t]{0.42\textwidth}
        \centering
        \includegraphics[width=\linewidth]{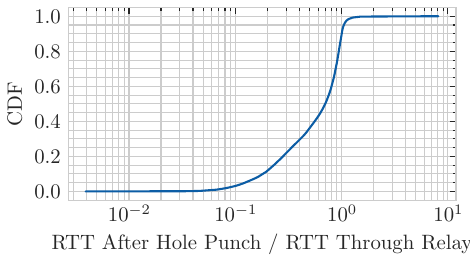}
        \caption{}
        \label{fig:latency-impact}
    \end{subfigure}%
    ~ 
    \begin{subfigure}[t]{0.5\textwidth}
        \centering
        \includegraphics[width=\linewidth]{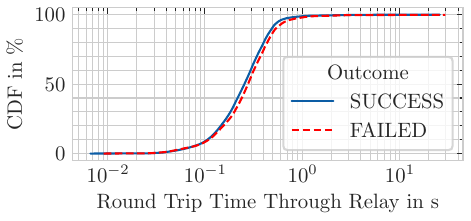}
        \caption{}
        \label{fig:rtt-dependence}
    \end{subfigure}%
    \caption{(a) The direct RTT as a fraction of the RTT through the relay. (a) RTT distributions for hole punches with the outcomes \texttt{SUCCESS} and \texttt{FAILED}. The sample sizes are $897,747$ and $684,376$ respectively.}
\end{figure*}

A critical requirement for a decentralized relay system is independence from the specific characteristics of the chosen relay, as stated in H2. We first test this hypothesis by investigating the influence of the round-trip time (RTT) through the relay on the hole punch outcome. Figure~\ref{fig:rtt-dependence} shows two Cumulative Distribution Functions (CDFs) of the round trip times through the relay to the remote peer in cases of successful (solid blue line) and failed hole punches (dashed red line). The graph indicates a weak negative correlation: failed hole punches exhibit slightly higher RTTs, suggesting that greater relay latency marginally decreases the probability of a successful traversal. However, the impact is small as both CDFs follow a similar distribution.



To further validate our hypothesis of relay independence (H2), we analyze whether the relay's location along the network path between the two peers impacts the success rate. 
The data reported by clients includes RTTs to the remote peer via the relay and separate RTT measurements to the relay itself.
This allows us to investigate how the location of the relay along the path to the remote peer influences the success rate. If we define the RTT through the relay to the remote to be $100\%$ of the distance, then the RTT to the relay will be a fraction of that RTT. For example, if the RTT to the remote peer through the relay is $1\;\text{s}$ and the RTT to the relay is $700\;\text{ms}$, we define that the relay is $70\%$ away from the client to the remote peer. Figure~\ref{fig:relay-path-location-dependence} shows the success rate (in blue) as a function of the relay's path location, binned in $5\%$ increments, to the remote. Further, the graph shows in red the number of hole punch results where the relay was at that specific path location. The graph shows that the success rate is largely independent of the path location. Also, the majority of relay nodes fall right in the middle from us to the remote peer with a slight skew of the distribution towards the local client.

That the distribution is skewed towards the client is expected as peers by default request reservations with two relays. Clients then dial remote peers via all available relay addresses. It is natural that the connection to the closer relay and thus to the remote peer through it succeeds first.




Taken together, the findings show only a minor impact from high RTTs and no discernible impact from relay path location, provide compelling evidence for H2. This confirms that DCUtR's performance is robustly independent of the facilitating relay's network properties, a crucial feature for a permissionless P2P system.

\subsubsection{H3: Effective Synchronization}
\label{sec:effective-synchronization}

\begin{figure*}[t!]
    \centering
    \begin{subfigure}[t]{0.5\textwidth}
        \centering
        \includegraphics[width=\linewidth]{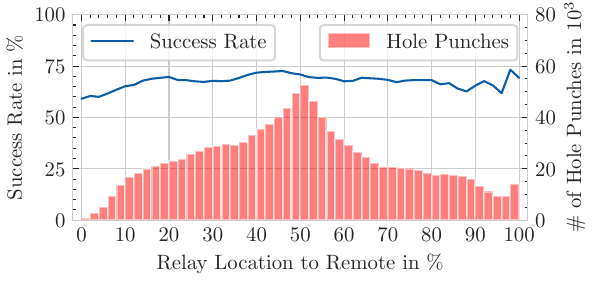}
        \caption{}
        \label{fig:relay-path-location-dependence}
    \end{subfigure}%
    ~ 
    \begin{subfigure}[t]{0.4\textwidth}
        \centering
        \includegraphics[width=\linewidth]{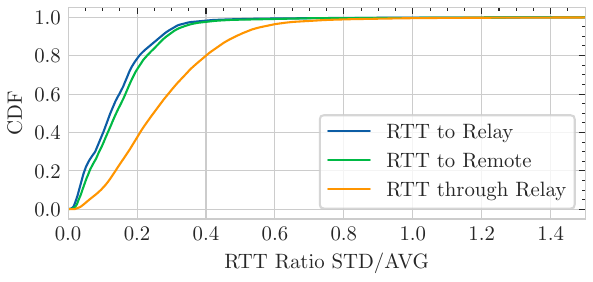}
        \caption{}
        \label{fig:rtt-accuracy}
    \end{subfigure}%
    \caption{(a) Success rate dependence on relay location along the path. (b) Ratio of the mean over the standard deviation of the RTT measurements}
\end{figure*}

The protocol's synchronized dialing mechanism is central to its design. We now test two related hypotheses: that the synchronization is highly efficient (H3a) and that it is effective enough to make the protocol transport-agnostic (H3b). To validate the underlying mechanism of our deterministic synchronization hypotheses, it is essential to first establish the accuracy of the RTT measurements that drive the protocol. Effective synchronization requires that these measurements are sufficiently precise to coordinate the simultaneous connection attempts where packets have already departed from the respective routers of the peers before the arrival of packets from the other peer. Figure~\ref{fig:rtt-accuracy} presents all RTT measurements collected during the measurement campaign. Each RTT measurement comprises up to ten ping samples. The dataset is categorized into three distinct RTT measurement scenarios: (1) from the client to the relay, (2) from the client to the remote peer via the relay, and (3) from the client directly to the remote peer following successful hole punching. The results indicate that, for the critical RTT measurements between the client and the remote peer through the relay, in over $90\%$ of instances, the standard deviation of latency measurements constitutes only half of the average RTT. This demonstrates a considerable synchronization margin. As expected, latency measurements to the relay and to the remote peers exhibit similar distributions, with RTT variations generally smaller than those observed in the measurements involving the relay path to the remote peer.


\pb{Attempts.} Having established a notion of RTT measurement accuracies we go ahead and test our hypothesis (H3a). An examination of successful hole punches in Figure~\ref{fig:attempts-success} reveals that the vast majority, $97.6\%$, succeeded on the very first attempt. Only a small minority of $2.4\%$ of successful hole punches required one of the subsequent attempts. This observation carries immediate and significant implications for protocol optimization. If subsequent attempts yield only marginal gains, the DCUtR protocol could be refined to reduce the number of retries or change its strategy. This crucial finding can directly feed back into the protocol design. We discuss protocol improvements in Section~\ref{sec:optimizations}.





\pb{Transport Protocol Dependence.} We now address one of our central hypotheses (H3b), which challenges the conventional wisdom that UDP-based hole punching is inherently superior to TCP-based methods. This belief stems from several fundamental differences in how TCP and UDP operate and how NATs and firewalls process their traffic. We discuss the common arguments in Section~\ref{sec:discussion}. To test this claim of transport agnosticism, we analyze the success rates for hole punches that were explicitly restricted to either TCP or QUIC.



\begin{wrapfigure}{r}{.5\textwidth}
    \centering
    \vspace{-0.6cm}
    \begin{subfigure}[t]{0.25\textwidth}
        \centering
        \includegraphics[width=\linewidth]{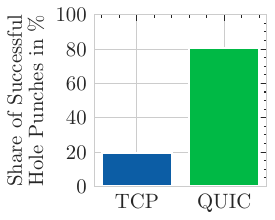}
        \caption{}
        \label{fig:transport-success-share}
    \end{subfigure}
    \begin{subfigure}[t]{0.24\textwidth}
        \centering
        \includegraphics[width=\linewidth]{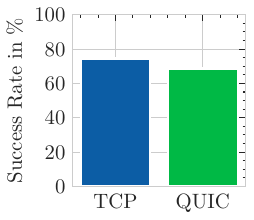}
        \caption{}
        \label{fig:transport-success-rate}
    \end{subfigure}
    \vspace{-0.1cm}
    \caption{(a) If the hole punch was successful, which transport was the final connection using (b) If the attempts were restricted to using a single transport, what are the respective success rates.}
    \vspace{-0.3cm}
\end{wrapfigure}

Figure~\ref{fig:transport-success-share} shows that if we don't restrict the transport protocol that's being used to do the hole punch, the final connection is in the vast majority $\sim 80\%$ of cases using QUIC. However, this only shows that connection establishment in QUIC is faster than in TCP and does not tell anything about the success rate. This hypothesis is strongly supported by recent, independent research by Liang et al.~\cite{liang2024}. If we look at the data where we had applied a protocol filter and restricted the hole punch to a certain transport protocol we find in Figure~\ref{fig:transport-success-rate} that the success rates are both close to $\sim70\%$, strongly validating (H3b). This finding demonstrates that DCUtR's synchronization mechanism effectively mitigates the traditional challenges of TCP traversal, making it a transport-agnostic protocol.

\subsubsection{H4: Effectiveness of the Connection Reversal Optimization}
\label{sec:optimization-effectiveness}

The DCUtR protocol includes Connection Reversal as a fast-path optimization. To test its real-world effectiveness (H4), we analyze the distribution of outcomes for peers that report an active port mapping. As noted in Section~\ref{sec:dcutr}, libp2p attempts to register port mappings with routers using UPnP or PMP. Hole punch result data points from our fleet of clients also comprise information about active port mappings in their local network. However, while these mappings are considered advisory due to potential router dishonesty or unannounced expirations, their presence is expected to influence hole punching outcomes. Specifically, a higher share of \texttt{CONNECTION\_REVERSAL} outcomes would be anticipated when an active port mapping is reported, as this allows a direct connection without the need for a full hole punch.

\begin{figure*}[t]
    \centering
    \begin{subfigure}[t]{0.4\textwidth}
        \centering
        \includegraphics[width=\linewidth]{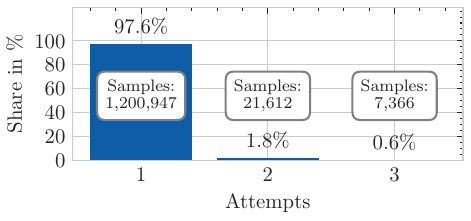}
        \caption{}
        \label{fig:attempts-success}
    \end{subfigure}%
    \hspace{0.05\textwidth} 
    \begin{subfigure}[t]{0.45\textwidth}
        \includegraphics[width=\linewidth]{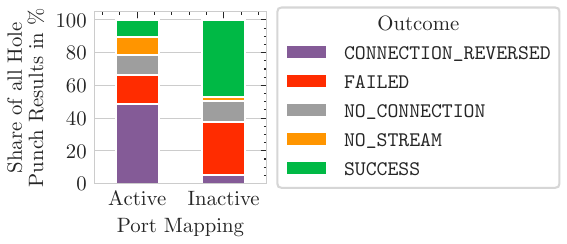}
        \caption{}
        \label{fig:port-mappings-outcomes}
    \end{subfigure}%
    \caption{(a) If successful, with which attempt succeeded the hole punch. (b) Influence of active port mappings on the hole punch outcome.}
\end{figure*}

Figure~\ref{fig:port-mappings-outcomes} shows the distribution of outcomes in the cases where a port mapping was active and inactive respectively. One can clearly see the significantly higher share of \texttt{CONNECTION\_REVERSAL} outcomes in the case of active port mappings on the clients' side of the connection. This observation empirically validates the importance of the connection reversal technique within the DCUtR protocol. If a direct connection can be established via a pre-existing port mapping, it is highly advantageous to prioritize this method over initiating a full hole punch. Connection reversal avoids the inherent complexities, timing sensitivities, and resource consumption associated with the hole punching procedure, making it a more efficient and reliable path to direct connectivity when applicable. This empirical validation of the Connection Reversal mechanism confirms H4 and underscores the value of including such preliminary checks to avoid the overhead of the full hole-punching procedure when possible.

%% file: sections/6.discusssion.tex
\section{Discussion}
\label{sec:discussion}

Our empirical evaluation has validated the core hypotheses about DCUtR's performance. The confirmation of relay independence (H2) is paramount for a permissionless P2P network, as it proves that the system does not rely on a privileged or well-positioned set of relays for its core function. This robustness is critical in environments where relay selection is random and relay quality is not guaranteed. Furthermore, the establishment of a contemporary baseline success rate of approximately $70\%$ across a large fraction of the network over an extended period (H1) surpasses the scale and recency of prior work and provides a crucial benchmark for future optimizations and comparisons. Earlier findings from Halkes et al.~\cite{Halkes2011-og} in 2011 found that ``approximately $64\%$ of all peers are behind a NAT box or firewall which should allow hole punching to work, and more than $80\%$ of hole punching attempts between these peers succeed.'' To the best of our knowledge, only Guha et al.~\cite{Guha2005} reported ``in the wild'' hole punching success rates. They report an ``88\% average success rate for TCP connection establishment with NATs in the wild.'' However, their measurement comprises just 93 home NATs, which provides substantially less statistical power and representativeness than our considerably larger dataset. Moreover, all of the studies we identified rely on data that is considerably outdated, which may not accurately reflect the current internet landscape and underscore the lack of recent analyses.

\pb{Transport Protocol Dependence.} The validation of H4 (Transport Agnosticism) warrants a deeper discussion. The theoretical challenges of TCP hole punching around the strict three-way handshake and susceptibility to \texttt{RST} packets from stateful firewalls are well-documented~\cite{rfc5128,Guha2005}. Further, the standard Berkley sockets API does not inherently support the simultaneous open and listening on the same port required for TCP hole punching. While \texttt{SO\_REUSEADDR} and \texttt{SO\_REUSEPORT} options can facilitate this, their improper use may violate TCP standards and introduce complex, hard-to-debug issues. Also, the repetitive SYN transmissions inherent in traditional TCP hole punching simultaneous open attempts can be misconstrued as SYN flood attacks by network security systems, leading to blocked connections or temporary IP blacklisting. Moreover, TCP's stateful nature demands the allocation of router state table resources. Persistent, unsuccessful TCP hole punching attempts could exhaust these limited resources, potentially degrading network performance, leading to denial-of-service conditions in constrained network environments or prohibiting subsequent retries. 

The measurement campaign's findings regarding transport protocol dependence present a significant empirical counterpoint to this conventional understanding. In our data, the success rates for both TCP and QUIC were both around $70\%$. This directly contradicts the ``common belief that UDP hole punching is easier than TCP'' challenging long-held assumptions. The reconciliation of this unintuitive finding with the detailed theoretical difficulties of TCP hole punching suggests DCUtR's synchronization effectiveness.  Unsolicited SYN packets would only reach the other peers' NAT if the error of the RTT measurement through the relay is higher than the one way latency between the two NATs being traversed which we have confirmed to not being the case with the RTT accuracy measurements. These showed that the standard deviation is less than half of the RTT in the vast majority of cases. The synchronization accuracy means, the problem space for DCUtR largely becomes deterministic instead of stochastic. This insight drives the following optimizations in Section~\ref{sec:optimizations}.

\pb{Limitations.} As detailed in Section~\ref{sec:network-identification}, our analysis excludes approximately $29\%$ of collected data points, primarily those where prerequisite protocols failed to establish a relayed connection or identify a public address. Therefore, our reported success rates reflect the performance of the DCUtR hole-punching mechanism itself, conditional on the success of these earlier stages, rather than the end-to-end success rate of establishing a direct connection from scratch. This identification failure rate underscores the multi-faceted challenges of establishing P2P connectivity in the wild.

\section{Optimizations \& Future Work}
\label{sec:optimizations}

The persistent $\sim30\%$ failure rate represents the next frontier for research. These failures can stem from a myriad of factors, including transient network conditions, misconfigurations, overly restrictive firewalls, or the prevalence of Symmetric NATs. In this Section, we propose a multi-tiered roadmap to universal connectivity. The viability of this roadmap is underscored by the central finding of transport-agnosticism of DCUtR as a result of the effective RTT-based synchronization.

\subsection{Probabilistic Traversal via the Birthday Paradox}
\label{sec:optimization:symmetric-nat}

The $\sim30\%$ failure rate, despite the protocol's general efficacy (H1), is likely due in part to challenging NAT configurations such as Endpoint-Dependent Mappings. To address this class of failures, which cannot be solved by simple synchronization, we propose exploring techniques based on the birthday paradox, a known method for traversing EDM NATs by creating a high probability of a port collision. Halkes et al.~\cite{Halkes2011-og} observed in 2011 that approximately $11\%$ of their monitored peer cohort were behind EDM NAT devices. As discussed in Section~\ref{sec:network-address-translation}, these devices make it impossible to predict the assigned port when establishing connections to other peers in the network, rendering the port information exchanged via the relay in the DCUtR protocol ineffective. 

To address this challenge we can exploit the statistical properties of port allocation to create collisions within the $2^{16}$ port search space that would otherwise be computationally intractable. For instance, if the peer behind the EDM NAT opens 256 ports simultaneously while the connecting peer probes 256 randomly selected ports (representing $0.4\%$ of the total search space), the probability of establishing a successful connection reaches $64\%$. When the search space coverage increases to $3.1\%$ (2048 ports), the success probability rises to $99.9\%$. However, when both peers operate behind EDM NATs, the success probability drops significantly to $0.01\%$ when probing 2048 ports on one side and 256 on the other~\cite{Anderson_2020}.
Based on the EDM NAT prevalence reported by Halkes et al.~\cite{Halkes2011-og}, we expect to encounter approximately $19.6\%$ mixed EDM/EIM peer combinations and $1.2\%$ combinations where both peers are behind EDM NATs. If birthday paradox techniques achieve a $64\%$ success rate for the $19.6\%$ of mixed connection attempts, the overall DCUtR success rate could improve by $12.5\%$.
This analysis assumes conservative values for port probing and opening operations, as NAT devices maintain limited active session capacities and may interpret intensive port scanning as malicious behavior, potentially blocking the peer's IP address for extended periods. Without these constraints, success rates could be substantially higher.

The data from Halkes et al.~\cite{Halkes2011-og} presents limitations due to its age (2011) and relatively small sample size (fewer than 2000 peers). Two conflicting trends have emerged since their study: \one the expected reduction in EDM NAT deployment as vendors increasingly comply with the BEHAVE RFC~\cite{rfc4787}, and \two the proliferation of Carrier-Grade NAT (CGNAT) deployments, which employed EDM translation in $40\%$ of cases as of 2016~\cite{Richter2016}. Future empirical studies are necessary to establish more accurate contemporary estimates of expected success rate improvements.

\subsection{Protocol Optimizations}
\label{sec:optimization:protocol}

\pb{Refining RTT Calculation for Asymmetric Routes.} A powerful optimization would be if users chose a proxy known to have advantageous NAT properties (e.g., a full-cone NAT). This strategy, however, introduces an additional network hop, making the timing of the hole punch even more critical. The current DCUtR protocol utilizes Round Trip Time (RTT) measurements between \textbf{peers} to coordinate hole punching attempts. However, it is the RTT between the respective NAT devices, rather than the peers themselves, that is truly pertinent. Although our measurements indicate that peer-to-peer RTT often serves as a reasonable approximation, this approach may be insufficient in more complex network topologies. In such cases, the resulting discrepancy can cause the \texttt{SYNC} packet to arrive outside the optimal time window, thereby increasing the likelihood of connection failures. Therefore, we propose an enhancement where the listener includes its observed  $\text{RTT}_{\text{Listener - NAT}}$ to its NAT (or an equivalent near-edge network element) within its \texttt{CONNECT} message. The initiator, upon receiving this information, can then compute a more accurate wait time $T_{\text{wait}}$ for sending its \texttt{SYNC} packet by also incorporating the $\text{RTT}_{\text{Initiator - NAT}}$ to its NAT. The refined calculation for the protocol's RTT then reads as follows: $T_{\text{wait}} = 1/2( \text{RTT}_{\text{Listener - Initiator}} + \text{RTT}_{\text{Listener - NAT}} -\text{RTT}_{\text{Initiator - NAT}} )$.
This approach aims to better estimate the one-way traversal time to peer B's NAT, thus further improving the timing precision of the \texttt{SYNC} message and increasing hole punching success rates in networks with asymmetric path latencies. 

\pb{Alternating Roles on Connection Retries.} The validation of H4 revealed that while the first attempt is highly successful, subsequent retries yield diminishing returns. This suggests that simply repeating the same action is suboptimal. We therefore propose an alternative retry strategy for subsequent attempts to increase the success rate. In the case of QUIC hole punching, upon receiving the SYNC message from the initiator, the listener immediately dials back. The initiator on the other hand starts to send UDP packets filled with random bytes to the listener upon expiry of the 1/2 RTT timer to prime its NAT. This will result in a QUIC connection where the listener is the client and initiator is the server. If we assume that the initiator is behind an EDM NAT but the listener is not, the listener's client hello will not be able to traverse the initators NAT because of the unpredictable nature of the port mappings. We propose to alternate the roles in subsequent attempts. The above scenario would succeed on the second try, given that the initiator is not behind a symmetric NAT. Contrary to a TCP simultaneous open, having both parties sending client hellos might result in two distinct QUIC connections which can also be acceptable in some scenarios.

\pb{Proactive NAT Priming to Prevent Denylisting.} As noted in~\cite{rfc5128}, some NATs or firewalls may temporarily denylist external IPs if inbound UDP packets arrive before any outbound traffic is sent to those addresses. In DCUtR, this can occur if the initiator’s hole punching packets reach the listener’s NAT before the listener initiates traffic, potentially causing the initiator’s address to be blocked. To mitigate this, we propose a more cautious NAT priming strategy. Instead of sending random UDP payloads that might reach the initiator’s NAT, the listener should send low Time-To-Live (TTL) packets (e.g., TTL=3) to the initiator’s advertised addresses. This limits propagation while priming the listener's NAT. Similarly, once the initiator receives the listener’s \texttt{CONNECT} message, it should send low-TTL packets to the listener until its \texttt{SYNC} timer fires. This staggered priming helps ensure both NATs have outbound mappings before hole punching begins, reducing the risk of denylisting.

\subsection{Ecosystem Strategies}
\label{sec:optimization:ecosystem}



Long-term success also depends on improving the network environment itself through a combination of standards advocacy and better use of existing mechanisms. This echoes a long-standing hope in the community like in 2011 by Halkes et al.~\cite{Halkes2011-og}, that broader vendor compliance with standards like IETF's BEHAVE~\cite{rfc4787} would reduce the prevalence of such ``hard'' NATs. In parallel, our findings confirm that the Connection Reversal optimization is highly effective when a port mapping is available (H4). Therefore, client software can play a more active role by guiding users to enable UPnP/PMP on their routers, explaining the direct connectivity benefits while acknowledging potential security trade-offs.

%% file: sections/7.related-work.tex
\section{Related Work}
\label{sec:related-work}

NAT hole punching, or NAT traversal, has been a recognized technique for enabling direct peer-to-peer communication across Network Address Translators for over two decades. The concept was initially introduced by Dan Kegel in 1999~\cite{kegel_1999}, who outlined a NAT traversal protocol utilizing UDP packets, primarily motivated by the requirements of peer-to-peer gaming applications. One of the foundational publications in this domain is by Ford et al.~\cite{Ford2005PeertoPeerCA}, which describes early approaches to peer-to-peer communication across NATs. Subsequent research has explored various facets of NAT traversal. Guha and Francis~\cite{Guha2005} provided an empirical characterization and measurement of TCP traversal through NATs and firewalls, offering insights into the challenges posed by TCP's connection-oriented nature. Similarly, Halkes and Pouwelse~\cite{Halkes2011-og} investigated UDP NAT and firewall puncturing in real-world scenarios, contributing to the understanding of UDP's behavior in NAT environments. The Interactive Connectivity Establishment (ICE) protocol~\cite{rfc8445, rfc5245} unifies several NAT traversal techniques, including Session Traversal Utilities for NAT (STUN)~\cite{rfc8489} and Traversal Using Relays around NAT (TURN)~\cite{rfc8656}, to provide a comprehensive framework for establishing peer-to-peer connections in complex network topologies. ICE is widely adopted in real-time communication protocols such as WebRTC~\cite{rfc8825}. Further studies have delved into specific aspects and challenges of NAT traversal. Ford et al.~\cite{Ford2005PeertoPeerCA} provided a comprehensive overview of the state of peer-to-peer communication across NATs, detailing various techniques and their limitations. Richter et al.~\cite{Richter2016} conducted a multi-perspective analysis of Carrier-Grade NAT (CGNAT) deployment, highlighting its prevalence and impact on network behavior. Other works have explored edge-case integration into established NAT traversal techniques~\cite{Keller2022}, knowledge-based NAT-traversal for home networks~\cite{muller2008}, and the revisiting of NAT hole punching strategies~\cite{Maier2011}. Additionally, research has examined peer NAT proxies for peer-to-peer games~\cite{Seah2009}. Liang et al.~\cite{liang2024} show that QUIC-based hole punching is theoretically 0.5 RTT faster than TCP-based hole punching due to its integrated 1-RTT handshake compared to TCP's 3-way handshake followed by a separate TLS handshake.


%% file: sections/8.conclusion.tex
\section{Conclusion}
\label{sec:conclusion}

This paper presented a comprehensive large-scale empirical evaluation of the DCUtR protocol, validating our core hypotheses. The study establishes a crucial, contemporary baseline success rate of approximately $70\%$ for the hole punching stage (H1), demonstrating the viability of decentralized NAT traversal. We confirmed that success is robustly independent of relay network characteristics (H2), a key requirement for permissionless systems. Furthermore, we showed the protocol's synchronization is highly efficient, with successes predominantly occurring on the first attempt (H3a), and empirically challenged long-standing ``tribal knowledge'' by demonstrating comparable success rates for both TCP and QUIC (H3b). Finally, our analysis confirmed the effectiveness of key protocol optimizations like Connection Reversal (H4). The observed latency reduction post-hole punch underscores the protocol's ability to enable a broader range of delay-sensitive decentralized applications. We contribute a unique, large-scale, and CC BY-SA licensed dataset comprising over 4.4 million data points from a globally distributed cohort in over 85k individual networks, fostering reproducibility and opening new avenues for future research.

The insights gained from this campaign pave the way for more robust, resilient, and performant decentralized peer-to-peer applications building on top of libp2p. By reducing reliance on centralized intermediaries, DCUtR contributes to the foundational goals of the decentralized web: mitigating single points of failure and enhancing self-sovereign control over data. The identified areas for protocol optimization, such as role alternation on retries, refined RTT calculation, proactive NAT priming, and the birthday paradox exploitation offer promising avenues for further increasing direct connectivity and future research. Overall, DCUtR can not fully replace traditional STUN/TURN services yet. However, it can reduce the load on STUN/TURN servers of about 70\% today and with in combination with using relays (such as VPN or SSH Gateways) replace them entirely.

%% file: sections/9.appendix.tex
\clearpage
\section{Appendix}

\subsection{DCUtR Sequence Diagram}
\label{appendix:dcutr-sequence-diagram}

\begin{figure}[h]
    \centering
    \includegraphics[width=\textwidth]{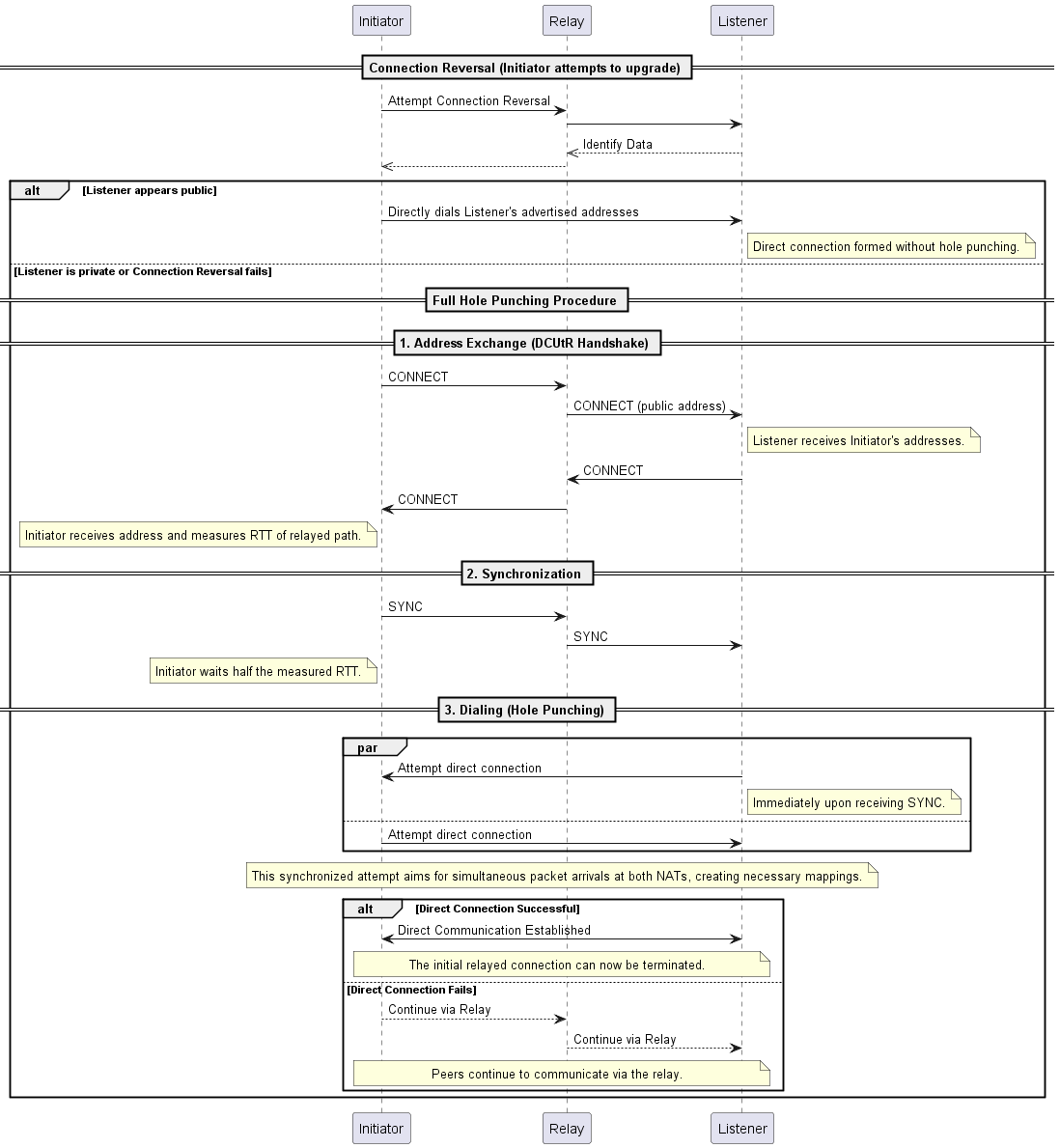}
    \caption{DCUtR Protocol Sequence Diagram}
    \label{fig:dcutr-sequence-diagram}
\end{figure}

\subsection{Dataset}
\label{appendix:dataset}

\begin{figure}[h]
    \centering
    \includegraphics[width=\textwidth]{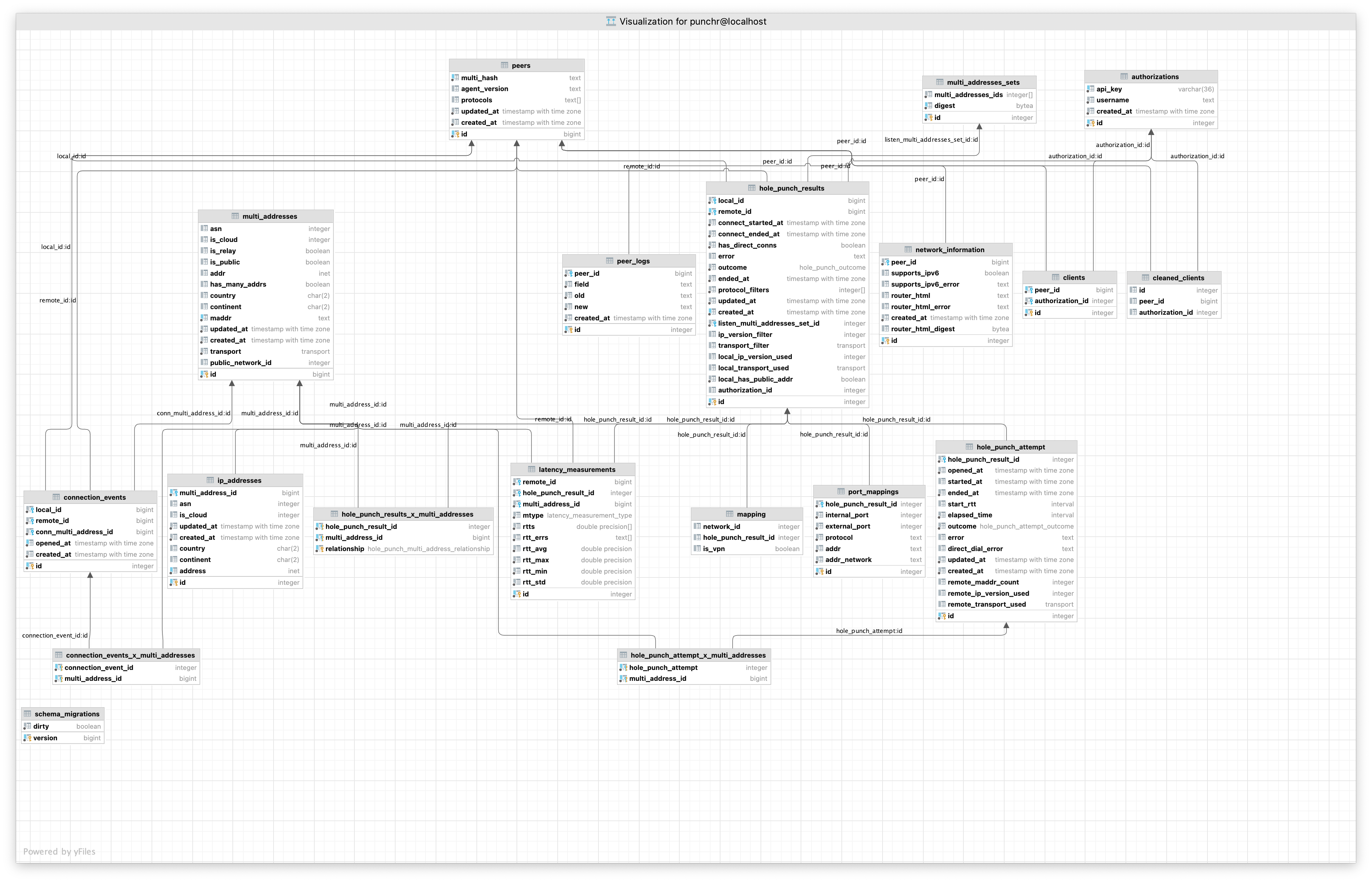}
    \caption{Postgres database UML diagram}
    \label{fig:psql-db-uml}
\end{figure}

In this section, we will provide a detailed overview of the dataset used in our analysis. This includes information on the time scope of the data, as well as instructions on how to download the dataset and set up a database for storage. Additionally, we will describe the process for restoring the data, and how we obtained the database dump.

\subsubsection{Peers/Multiaddresses}

We collect agent version, supported protocols, and Multihashes of all peers that interact with our measurement infrastructure. Similarly, when we get hand of a multi address we extract the underlying IP address, corresponding Geolocation (continent, country), if it’s a public address or not, if it’s a relayed address or not, if the IP address belongs to a known datacenter, and the corresponding Autonomous System Number.

\subsubsection{Connection Events}

As discussed in Section~\ref{sec:measurement}, the honeypot component tracks inbound connections from DCUtR capable peers. We call these events ``Connection Events''. A Connection event consists of the following information:

\begin{itemize}
    \item \texttt{local\_id} -- The internal database peer ID of the honeypot
    \item \texttt{remote\_id} -- The internal database peer ID of the peer that connected to the honeypot
    \item \texttt{conn\_multi\_address\_id} -- The multiaddress of the connection to the honeypot
    \item \texttt{connection\_events\_x\_multi\_addresses} -- All advertised multi addresses of the remote peer. They strictly won't contain a public IP address that is not a relayed address. This set will be served to clients.
\end{itemize}

\subsubsection{Hole Punch Results/Attempts}

A ``Hole Punch Result'' is what gets reported back by the clients and can consist of multiple ``Hole Punch Attempts''. Both data points can have different outcomes that are listed in Table~\ref{tab:hpr-outcomes} and Table~\ref{tab:hpa-outcomes}.

\subsubsection{Latency Measurements}

Before clients attempt to hole punch a remote peer, the client measures several latencies. It measures the latencies to all relays that the remote peer claims to be reachable through. It also measures the latency to the remote peer through one of the relays. Finally, if the hole punch succeeds, the client also measures latency of the direct connection. The database field \texttt{mtype} is \texttt{TO\_RELAY}, \texttt{TO\_REMOTE\_THROUGH\_RELAY}, and \texttt{TO\_REMOTE\_AFTER\_HOLEPUNCH} respectively.

\subsubsection{Port Mappings}

For each hole punch result, we track any active port mappings that get reported back from AutoNAT. If a port mapping is in place it is more likely for an outcome of \texttt{CONNECTION\_REVERSED}. With this data we can test this hypothesis. Each port mapping consists of the following data:

\begin{itemize}
    \item The hole punch it refers to
    \item Internal Port of the client in the local network
    \item External Port of the router
    \item Transport used for port mapping
    \item External facing address
\end{itemize}

\subsubsection{Authorizations/Clients}

Each time someone signed up through our Google Form we generated a UUID API-Key and saved it alongside the email address (from the form) into this table. We asked participants to provide the API Key to their client installations (GUI or CLI). Every hole punch result reported from these clients contained the API key so that we can associate the result with the Google Form information.

Importantly, the Go-Client generated ten peer identities upon startup. Each peer ID would listen on a different port. For each hole punch we ‘round robin’ through the ten peers. This way we mitigated persistent port mappings to yield a high number of \texttt{CONNECTION\_REVERSED} results. When the client starts up it reports the identities of these ten clients to the server. This fills the `clients` table where you have a mapping between peer ID and authorization ID.

So, to map a hole punch result to an authorization, you’d need to map the \texttt{local\_id} field onto the \texttt{clients} table and then to the \texttt{authorizations} table.

